# Dynamical renormalization of the magnetic excitation spectrum via high-momentum nonlinear magnonics

C. Schönfeld[1], L. Feuerer[1], J. Bär[1], L. Dörfelt[1], M. Kerstingskötter[1], T. Dannegger[1], D. Wuhrer[1], W. Belzig[1], U. Nowak[1], A. Leitenstorfer[1], D. Juraschek[2,3] and D. Bossini[1]*

[1]Department of Physics and Center for Applied Photonics, University of Konstanz, D-78457 Konstanz, Germany

[2]School of Physics and Astronomy, Tel Aviv University, Tel Aviv 6997801, Israel.

[3]Department of Applied Physics and Science Education, Eindhoven University of Technology, Eindhoven, Netherlands.

*Corresponding author. Email: davide.bossini@uni-konstanz.de.

## Abstract

**Controlling macroscopic properties of quantum materials requires the ability to induce and manipulate excited states. The set of collective excitations of a solid is encoded in its dispersion relations. We find that the spectra of the low-momentum eigenmodes are renormalized by resonantly driving the high-momentum excitations. Our experimental data rule out laser-induced thermal processes as an origin of the renormalization. The photoinduced changes of the amplitudes and frequencies can be explained theoretically and numerically with a resonant light-scattering mechanism that couples high- and low-momentum eigenmodes in the dispersion relation across momentum space. While we demonstrate the renormalization of the magnetic spectrum in a quantum material, our experimental approach can be further generalized to lattice modes in semiconductors.**

A central challenge of modern condensed-matter physics concerns the optical manipulation of quantum materials at fundamental timescales, which are defined by the oscillation cycles of collective vibrational, electronic, and magnetic excitations. Several milestones, such as photoinducing metal-insulator transitions (*1–3*), the quench (*4, 5*) and even the reversal of the magnetic order parameters (*6*) have been reported. Of particular interest is the generation of coherent collective lattice (*7*) and magnetic (*8*) modes, as they enable imprinting a well-defined phase on a macroscopic ensemble without requiring laser-induced heating (*8*). Coherent collective excitations have been driven into nonlinear regimes, displaying coupling among different modes (*7, 9–14*). While the phase and amplitude of coherently driven modes have been widely manipulated (*15, 16*), even for high-momentum modes near the Brillouin zone edges ("high-momentum" in the following) (*17*), an arbitrary optical control of the spectrum – in particular of the frequency - of the eigenmodes is lacking. Establishing this effect means changing the dispersion relations at specific momentum, thus directly manipulating the intrinsic dynamical response of the medium to external stimuli. So far, optically activated couplings between collective modes have been limited to excitations at the zone center ("Gamma-point" in the following).

Here we report the renormalization of the spectrum of Gamma-point magnetic eigenmodes in the quantum material α-Fe₂O₃ (hematite) at room temperature. Optically exciting pairs of high-momentum collective magnetic modes, i.e. magnons, in a resonant and selective manner results in a nonlinear coupling to two Gamma-point magnon modes. As a result, the amplitudes of the Gamma-point magnons are strongly enhanced and their eigenfrequencies renormalized, in comparison with an off-resonant excitation. We develop a formalism describing the resonant Raman scattering process, mediated by a nonlinear coupling of the modes across momentum space, which results in the amplification of the Gamma-point magnons. Supported by atomistic spin dynamics simulations, we ascribe the frequency renormalization to scattering among a variety of magnon modes throughout the dispersion relation.

## High-momentum magnons and experimental approach

Our material of choice is hematite, an antiferromagnet which displays a weak-ferromagnetic equilibrium state at room temperature. This is the initial state for all the experiments here reported. The magnetic eigenmodes of hematite relevant for our purposes are highlighted in the dispersion relation shown in Fig. 1**A**. The two Gamma-point modes, namely the quasi-ferromagnetic (q-FM) and quasi-antiferromagnetic (q-AFM) modes, constitute the magnetic resonances of hematite. The literature reports that the absorption spectrum of hematite in the mid-infrared range displays a peak, corresponding to the generation of high-momentum magnon pairs (*18*). This collective excitation is called two-magnon (2M) mode and it is driven in hematite by a spin-conserving electric dipole transition (see Supplementary Materials, "Materials" section). Theoretical descriptions of this process demonstrate the 2M electric dipole to be a function of solely the exchange interaction (*19–21*). Figure 1**B** displays the absorption spectrum of our specimen including a pronounced maximum at 45.5 THz induced by the 2M excitation (*18*). We perform a pump-probe experiment based on the scheme in Fig. 1**C**. Femtosecond laser pulses in the mid-infrared can excite hematite either resonantly ("ON" in Fig.1**B**) or off-resonantly ("OFF" in Fig.1**B**) with the 2M mode, as their central frequency (energy) is tunable from 37 to 53 THz (photon energies between 153 and 219 meV). The transient spin dynamics is detected via changes in the ellipticity of temporally delayed probe pulses, with a central wavelength of 1.2 μm (1 eV), i.e. in a transparent region of hematite. In our experimental geometry, second-order magneto-optical effects, such as magnetic linear birefringence, induce changes of the probe ellipticity (*22*). It follows that our scheme is sensitive to the in-plane components of the transient spin dynamics (*23*, *24*). The light source employed in our experiment is described in Ref. (*25*) and the details of the set-up are available in the Methods section of the Supplementary Materials. SQUID measurements (Supplementary Fig. S1) of our sample demonstrate that a magnetic field of 200 mT saturates the magnetization. Consequently, we always applied an in-plane field with this intensity (see Fig. 1**C**) to perform experiments under stroboscopic conditions.

## Renormalization of the spectrum

We begin our investigation by exciting the material off-resonantly. The central frequency of the pump beam is tuned to 37 THz and the photoinduced spin dynamics is represented by the yellow trace in Fig. 2**A**. Two oscillatory components are discernible. The Fourier transform of the time trace (Fig. 2**B**) exhibits maxima at frequencies of 167 GHz and 21.5 GHz. These features correspond to the q-AFM and q-FM modes, respectively (*22*, *26*). The identification of the oscillations in Fig. 2**A** with the q-AFM and q-FM modes is further corroborated by measurements performed as a function of the externally applied magnetic field (Supplementary Fig. S8). The observation of coherent Gamma-point magnons triggered by pump laser pulses tuned to the transparency region of the material is not surprising. These are the canonical experimental conditions to induce coherent modes via impulsive stimulated Raman scattering (ISRS), as already demonstrated both for phonons (*27*) and magnons (*16*) including the 2M mode (*17*, *28*).

However, when we tune the central frequency of the pump beam to be resonant with the 2M mode, while keeping the fluence constant, the situation changes dramatically. The green trace in Fig. 2**B** reveals that a resonant drive of the high-momentum modes significantly affects Gamma-point magnons. The amplitudes of both modes are vigorously enhanced and the frequencies are modified as well (Fourier transform in Fig. 2**B**). We highlight that, despite the remarkable volume of research concerning the optical activation of any coherent collective excitation (phonons, magnons, plasmons, polaritons) in solids, neither a modification of the spectrum as shown in Fig. 2**B** nor a coupling between high-momentum and Gamma-point modes at fundamental timescales has been reported. This evidence demands a systematic investigation of the photoinduced spin dynamics, by tuning the central frequency of the pump pulses. Fitting the time-domain data (Supplementary Fig. S9) with the procedure described in the Supplementary Materials ("Methods" section), we extract the dependence of the amplitudes (Fig. 2**C**) and frequencies (Fig. 2**D**) of the q-AFM and q-FM modes on the pump central frequency. Both modes are amplified (x5 for the q-AFM mode and x3.5 for the q-FM mode) by a resonant drive of the 2M mode. The eigenfrequencies are affected as well: the q-AFM blueshifts by 4% while the q-FM mode redshifts by 3%. These values are obtained setting the laser fluence to 6 mJ/cm$^2$, which we increased up to 100 mJ/cm$^2$ to measure the dependence of the effect on the excitation fluence. We emphasize that the Gamma-point magnons are not resonantly driven, as their eigenfrequencies are orders of magnitude below the spectral content of our mid-infrared pulses (shown in Supplementary Fig. S3). Moreover, the fluence dependences of the amplitudes display a linear trend for both resonant and off-resonant excitation (Supplementary Fig. S10 and S11). A resonant drive of the 2M mode also gives rise to a linear fluence dependence of the frequency shifts (Supplementary Fig. S10**B** and S11**B**). In contrast, pumping hematite off-resonantly the frequency of the Gamma-point modes is barely affected, despite the ever-increasing amplitude as a function of the fluence (Supplementary Fig. S10**A** and S11**A**). It follows that the renormalization of the spectrum cannot be interpreted in terms of a magnonic anharmonicity that arises from large-amplitude spin precessions. Instead, it is driven by a coupling between the high-momentum and Gamma-point magnons.

**Non-thermal origin of the spectral renormalization**

The experiments performed pumping the sample off-resonantly the sample demonstrate that Gamma-point magnons are driven via ISRS, which is a non-thermal mechanism (*8*). In the case of resonant pumping, light is absorbed by the 2M transition. We therefore consider the possibility that the renormalization of the spectrum of the Gamma-point magnons is driven by energy dissipation following optical absorption. To assess this hypothesis we devise an experiment, in which the fluence and energy per pulse are kept constant, while the average laser heating is varied. This idea is implemented by blocking a variable number of pump pulses, effectively changing the duty cycle of the modulation of the pump beam (Fig. 3**A**). Comparing the dataset obtained by changing the laser-induced thermal load of the sample by a factor four, we observe that the modifications of the frequency (Fig. 3**B**) and amplitude (Fig. 3**C**) are barely affected. If the pump fluence is set to 6 mJ/cm$^2$, the corresponding temperature increase induced by single laser pulses is estimated to be 0.24 K by adopting the specific heat and mass density of hematite from Ref. (*29*). Hence laser-induced heating, both average and single-pulse, cannot explain the spectral renormalization.

Additionally, we investigate the polarization dependence of the absorption of resonant mid-infrared laser pulses and of the spin dynamics (Supplementary Fig. S13). The absorption is isotropic in the plane of the sample, while the spin dynamics is not. In fact, it displays the typical polarization dependence observed in ISRS experiments. Selecting different polarization states of the pump beam reverses the phase of the magnetic oscillation (Supplementary Fig. S12-S13) (*8*).

**Discussion: Amplification of the Gamma-point modes**

In view of the non-thermal origin of the renormalisation of the Gamma-point modes and of the polarisation dependence we have demonstrated, the ISRS canonical formalism ([8], [16]) is extended to the resonant case. We propose that the Raman cross section of Gamma-point magnons has a resonance, originating from the high-momentum modes. This aspect constitutes the main novelty of our mechanism, which we name two-magnon resonant Raman scattering (2MRRS). The spin dynamics triggered by the 2M excitation can be described in terms of a spin-correlation function ([17], [30]), $C_{ij,\mathbf{k}} \sim \langle S_{i,\mathbf{k}} S_{j,-\mathbf{k}} \rangle$, where $i,j$ indicate ionic sites and $\mathbf{k} = \pi/a$ in our case. Because the electric dipole moment of the 2M excitation is given by $p_{i,\mathbf{k}} = \Pi_{ijk,\mathbf{k}} S_{j,\mathbf{k}} S_{k,-\mathbf{k}}$, where $\Pi_{ijk,\mathbf{k}}$ is the coupling coefficient, we can write the interaction potential for the absorption process as

$$H_{\text{int}} = \Pi_{ijk,\mathbf{k}} C_{ij,\mathbf{k}} E_k. \qquad (1)$$

The resonantly driven 2M mode then couples nonlinearly to the q-AFM and q-FM modes. This spin dynamics can be treated in the macrospin approximation,

$$H_{\text{nonlinear}} = \kappa_{ijk,\mathbf{k}} C_{ij,\mathbf{k}}^2 S_k + \kappa'_{ijkl,\mathbf{k}} C_{ij,\mathbf{k}}^2 S_k S_l \qquad (2)$$

where the first and second terms describe the couplings to the q-FM and q-AFM modes, respectively. These terms resemble the forms of the inverse Faraday ([8]) and the inverse Cotton-Mouton ([31]) effects, respectively, whose Raman-tensor components are resonantly enhanced by the 2M mode. Measurements performed with circularly polarised pump pulses demonstrate that the q-FM mode follows the helicity dependence of the inverse Faraday effect ([8]) (Supplementary Fig. S14), while the q-AFM mode shows a dependence on the linear polarization of light typical of the inverse Cotton-Mouton effect ([8], [31]) (Supplementary Fig. S12). Equation (2) reveals that the coupling of the 2M mode with the two Gamma-point modes differs. This fact is in line with our observations of quantitative differences in the amplification, amount and spectral response of the frequency shifts of the two zone-center modes in Figs. 2**C**-**D**. The details of the model are provided in Supplementary Note 1. If the frequency of the pump beam is tuned far away from the resonance, the terms provided in Eq. (2) become negligible and the conventional ISRS description is recovered. A graphical illustration of our mechanism and the corresponding energy diagrams in comparison with the canonical ISRS are shown in Fig. 4**A**-**B**.

**Discussion: Frequency renormalization of the Gamma-point modes**

Spin dynamics can be described in terms of the Landau-Lifshitz-Gilbert (LLG) equation

$$\frac{d}{dt}\vec{S}_i = -\frac{\gamma}{(1+\alpha^2)\mu_s}\vec{S}_i \times \left(\vec{H}_i + \alpha\vec{S}_i \times \vec{H}_i\right) \qquad (3)$$

where $\vec{S}_i$ represents a spin, $\mu_s$ is the magnetic moment, $\gamma$ is the electron gyromagnetic ration, $\alpha$ is the Gilbert damping, and $\vec{H}_i$ is the effective field acting on the $i$-th spin $\vec{S}_i$. A linearized version of this equation and the Hamiltonian in Supplementary Eq. (S13), treated under a set of approximations, are the foundations of linear spin wave theory. The key model assumptions are then: (i) the magnetic interactions are constant (see Supplementary Eq. (S13)); (ii) spins have constant magnitude; magnons are (iii) negligible in number and (iv) non-interacting with each other. The predictions of linear spin wave theory are expressed by the dispersion relation (Fig. 1**A**). However, the resonant pumping of hematite results in modifications of the spectra of Gamma-point magnons, which is not explainable within linear spin wave theory. Since we resonantly induced high-momentum modes with high-intensity laser pulses, we expect a non-negligible population of collective excitations. Therefore, we relax approximation (iii) going beyond linear spin wave theory. We follow the approach outlined by Kittel, to calculate the effects of a strong magnon thermal population on the spectrum of collective modes in a simple ferromagnet (*32*). In our case however, a non-thermal laser-induced population of high-momentum magnons in a weak-ferromagnet is considered. Working with a minimal model Hamiltonian, the problem can be treated analytically (see Supplementary Note 2). We find that the Gamma-point eigenfrequencies are indeed renormalized, but with opposite signs in comparison with the experimental evidence (Fig. 2**C**-**D**), i.e. the q-AFM(q-FM) mode is predicted to redshift(blueshift).

Evidently, the approximations made in the linear spin wave theory are too simplifying to describe our experiment correctly. Therefore, we take a different approach by employing atomistic spin dynamics numerical simulations for a material specific Hamiltonian which was parameterized before on an ab-initio level (*33*). We calculate the spin dynamics in the time domain by solving the complete Landau-Lifshitz-Gilbert equation of motion, which includes nonlinear regimes of spin dynamics. The ability to simulate the evolution of the magnetic ensemble triggered by the selective activation of exclusively high-momentum modes is key to the present problem. We technically implement this step by means of a fictitious magnetic field, staggered and alternating in space for each unit cell (see Supplementary Note 3). We focus on the eigenfrequencies of Gamma-point modes and compare the outcome of the experimental fluence dependence (Fig. 4**C**) with the simulations performed for different intensities of the fictitious magnetic field (Fig. 4**D**). Both figures reveal a blueshift(redshift) of the q-AFM(q-FM) mode eigenfrequency, consistent with the experiment. Strikingly, our numerical approach confirms the central role played by the generation of high-momentum magnons in the observed frequency renormalization. In fact, we perform simulations triggering spin dynamics by a conventional Zeeman torque, which resonantly excites the q-AFM mode. In this case (stars in Fig. 4**D**) the spectrum of the Gamma-point modes is still renormalized due to anharmonicity effects, but differently from the experiments: both modes blueshift and their energy separation decreases (stars in Fig. 4**D**). The comparison between different simulations in Fig. 4**D** is analogue to the fluence-dependent measurements performed by pumping the 2M mode both off-resonantly and resonantly.

We highlight that our numerical approach still assumes constant (i.e. not time or temperature dependent) model parameters (i.e. spin wave theory assumption (i)), while relaxing the other assumptions. It follows that it is not necessary to assume that the magnetic interactions are modified in the transient states. We thus ascribe the renormalization of magnons to a combination of photoinduced population of high-momentum modes and a subsequent chain of magnon-magnon scattering events with modes throughout the entire Brillouin zone. This scenario is reminiscent of a different physical phenomenon, namely the parametric excitations of Gamma-point magnons in ferromagnets. A comprehensive theory of this process demonstrates that the magnon frequencies are renormalized by the product of the induced magnon population and the magnon-magnon scattering rate (*34*). A similar theoretical proposition has been formulated also for antiferromagnets, to predict nonlinear spin dynamics induced by applying a voltage (*35*). However, generalizing these steady-state theoretical frameworks into a rigorous and complete treatment for the present case (optical excitation of high-momentum magnons, fundamental timescales, nonequilibrium states, coupling throughout the dispersion relation) is well beyond the scope of the current manuscript.

We have demonstrated a non-thermal renormalization of the spectra of Gamma-point magnons in a quantum material. Further developing the modification of the eigenfrequencies of Gamma-point modes can result in coherent optical generation of magnetic instabilities and thus magnetic phase transitions if the frequency of the soft mode can be pushed close to zero. Moreover, our experimental approach can be applied to lattice collective modes in widely available and technologically relevant semiconductors like silicon, germanium and silicon carbide, as pairs of high-momentum phonons form infrared-active effective electric dipoles also in these non-polar materials (*36*).

**Acknowledgments:** This work was supported by the Deutsche Forschungsgemeinschaft (DFG) through the SFB1432 (425217212, Project B07). D.B. acknowledges the support of the DFG program BO 5074/1-1. The authors thanks Michaela Lammel for the SQUID characterization of the samples. D.B. thanks Joe Barker and Steve Johnson for useful discussions. The authors thank Stephan Eggert, Christian Beschle and Alessandro Baserga for technical support.

**Funding:**

Deutsche Forschungsgemeinschaft (DFG) grant 425217212 (SFB1432 – Project B07)

Deutsche Forschungsgemeinschaft (DFG) grant BO 5074/1-1 (Emmy Noether)

**Author contributions:** D.B. conceived and coordinated the project. A.L. designed and established the experimental method. C.S., L.F. and J.B. performed the experiments and analysed the data under the coordination of D.B. D.W. developed the model of the dynamical renormalisation of the Hamiltonian under the supervision of W.B. D.M.J. developed the model for the 2MRRS mechanism. M.K. and L.D. performed the atomistic spin dynamics simulations under the supervision of T.D. and U.N. D.B. wrote the manuscript with contributions from all other authors.

**Competing interests:** Authors declare that they have no competing interests.

**Data and materials availability:** All data are available in the main text or the supplementary materials.

**Supplementary Materials**

Materials and Methods

Supplementary Notes 1-4

Figs. S1 to S14

References (*36–55*)

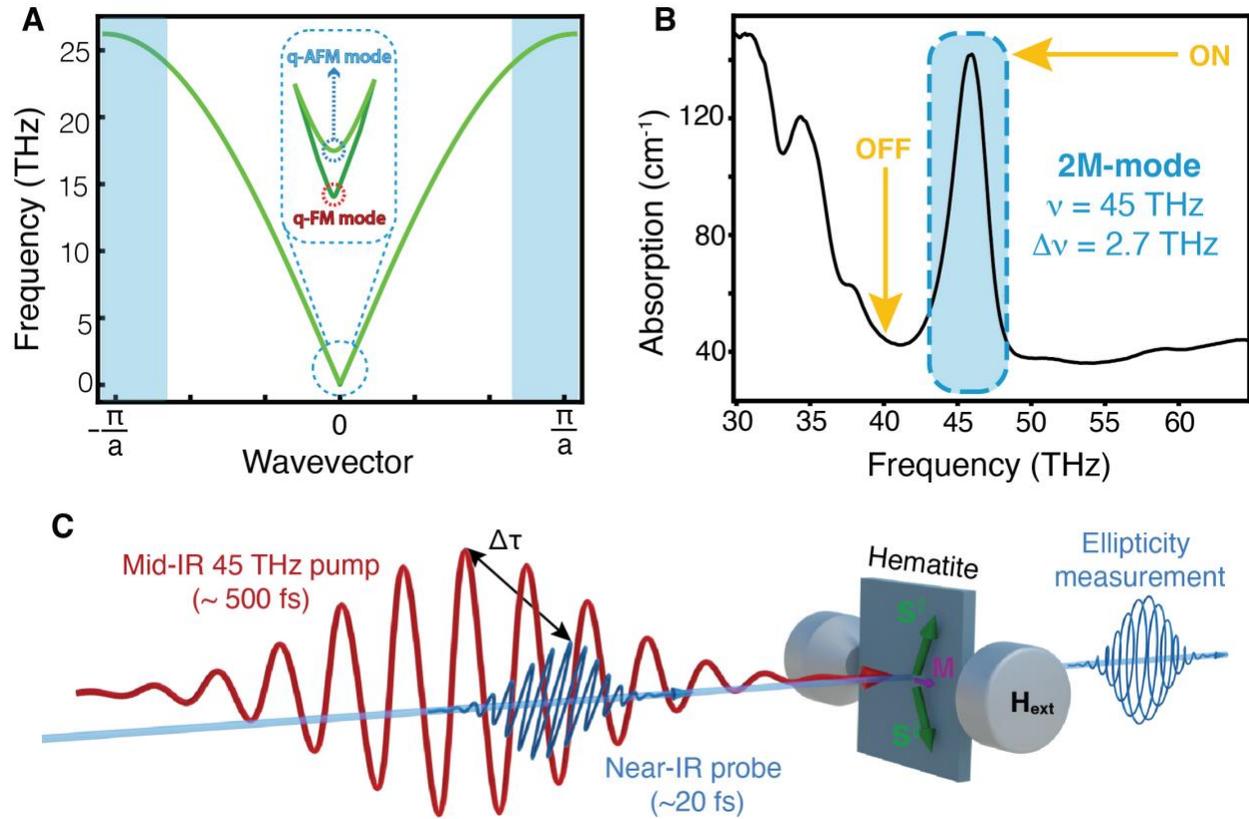

**Fig. 1. Hematite magnetic spectrum and concept of the experiment.** (**A**) Dispersion relation of the magnetic eigenmodes of hematite, as calculated within a two-sublattices approximation. The Gamma-point (q-FM and q-AFM) modes are shown in the inset, while the high-momentum modes that we activate optically are highlighted by the blue areas. (**B**) Room-temperature absorption spectrum of our specimen of hematite in the mid-infrared spectral range, as measured with an FTIR spectrometer. The maximum at ν = 45 THz (blue-dashed frame) with bandwidth Δν = 2.7 THz is attributed to the 2M mode (*18*). The frequency for resonant ("ON") and off-resonant ("OFF") pumping are indicated by yellow arrows. (**C**) Sketch of the experimental method. Femtosecond mid-infrared pump pulses trigger spin dynamics, which is detected by analyzing changes of the ellipticity of the near-infrared probe beam at a central wavelength of 1.2 μm. A magnetic field of 200 mT is applied in the plane of the sample by means of permanent magnets to saturate the magnetization (see Methods).

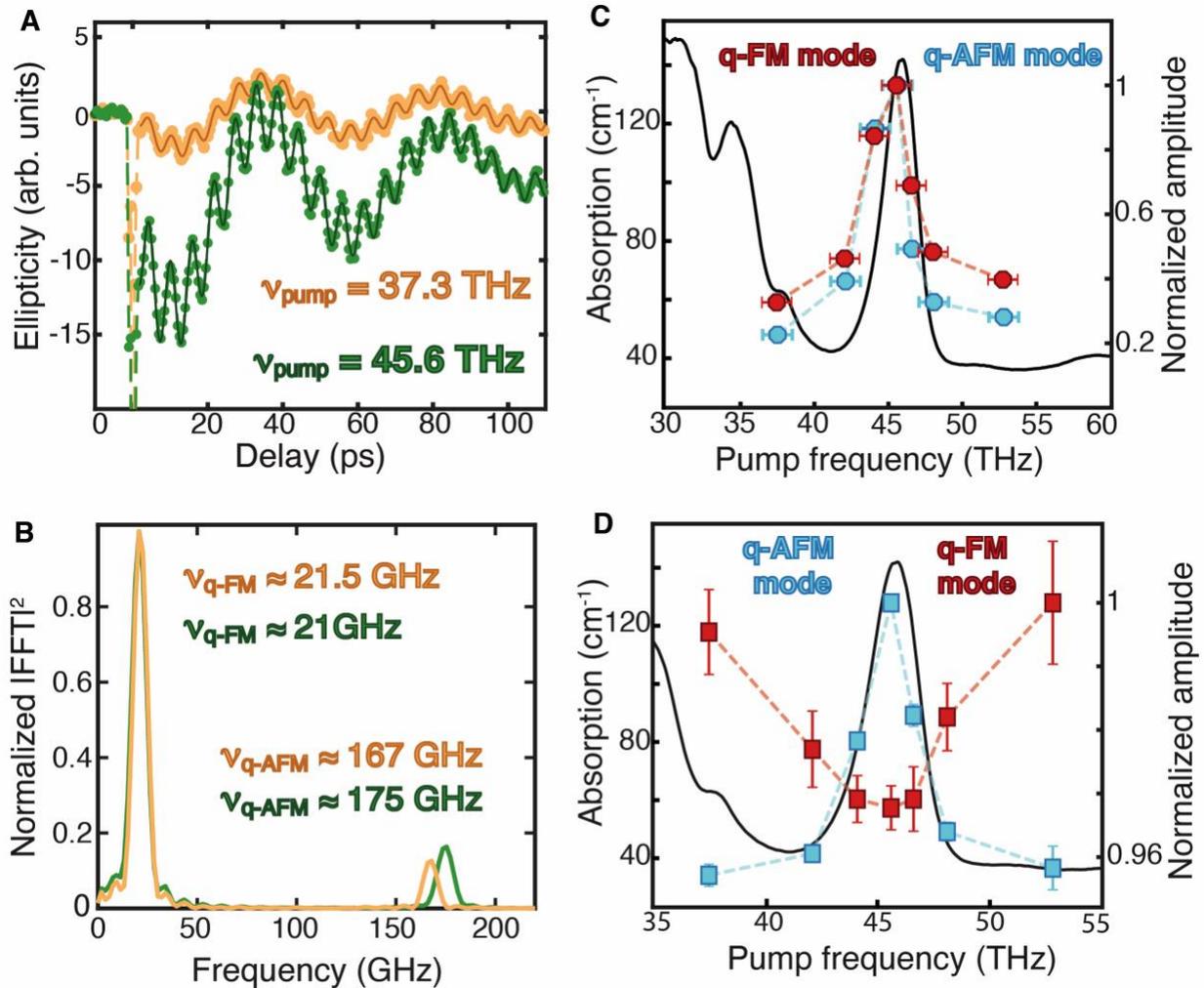

**Fig. 2. Renormalization of the spectrum.** (**A**) The spin dynamics photoinduced by pumping the 2M mode off-resonantly (yellow line) and resonantly (green line) is plotted. Both measurements were performed irradiating hematite with pump and probe beams linearly polarized along the direction parallel and orthogonal to the magnetic field (Fig. 1**C**), respectively. The fluence was set to 6 mJ/cm². The continuous lines represent least-square fits to the data (see Supplementary Materials – Section "Methods") (**B**) Spectra of the time traces in panel (**A**). The square modulus of the Fourier transform of the measurements in (**A**) is shown. The dependence of the amplitude (**C**) and of the frequency (**D**) of the Gamma-point modes on the pump beam frequency are shown. The data are normalized to the maximum value of each data set. In both panels, the absorption spectrum of hematite is displayed in the background (black line), including the 2M mode at 45 THz. Pump and probe beams were linearly polarized along the direction parallel and orthogonal to the magnetic field, respectively. The fluence was set to 6 mJ/cm². The error bars along the ordinate are defined as 95% confidence bounds of the fit. The error bars along the abscissa are given by the half width at half maximum of the intensity spectrum of the excitation.

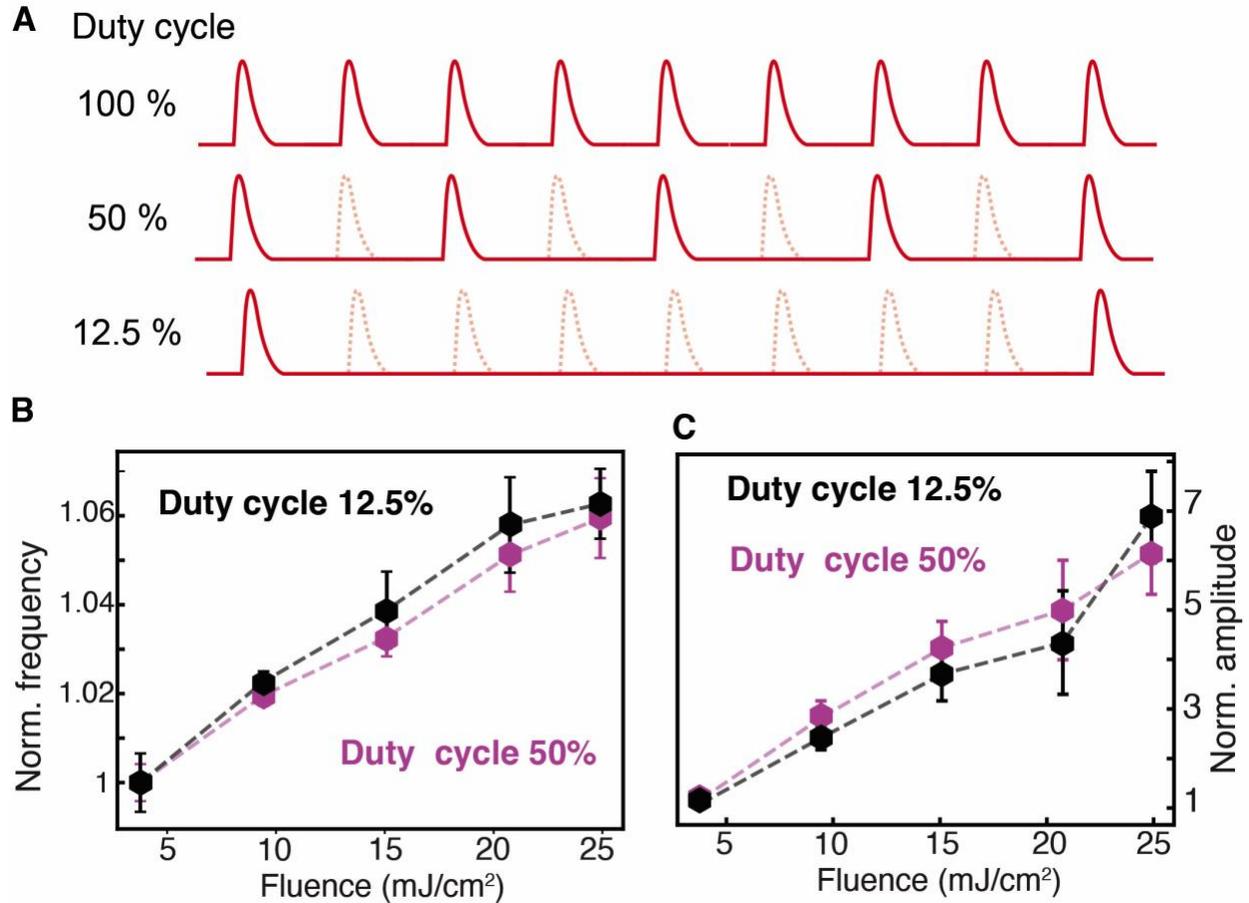

**Fig. 3. Non-thermal nature of the renormalization.** (**A**) Modulating the intensity of the pump beam by blocking a different amount of laser pulses changes the average power of illumination, i.e. the laser heating. Conversely the fluence and energy per pulse are kept constant. The fluence dependence of the frequency (**B**) and amplitude (**C**) of the q-AFM mode was studied for two different values of the duty cycle, namely 50% and 12.5%. The probe beam was linearly polarized along the direction orthogonal to the magnetic field, while the pump beam was orthogonally polarized to the probe. The excitation was resonant with the 2M mode. The measurements show that changing the accumulated heat deposition into the sample by a factor 4 significantly affects neither the frequency nor the amplitude of the q-AFM mode.

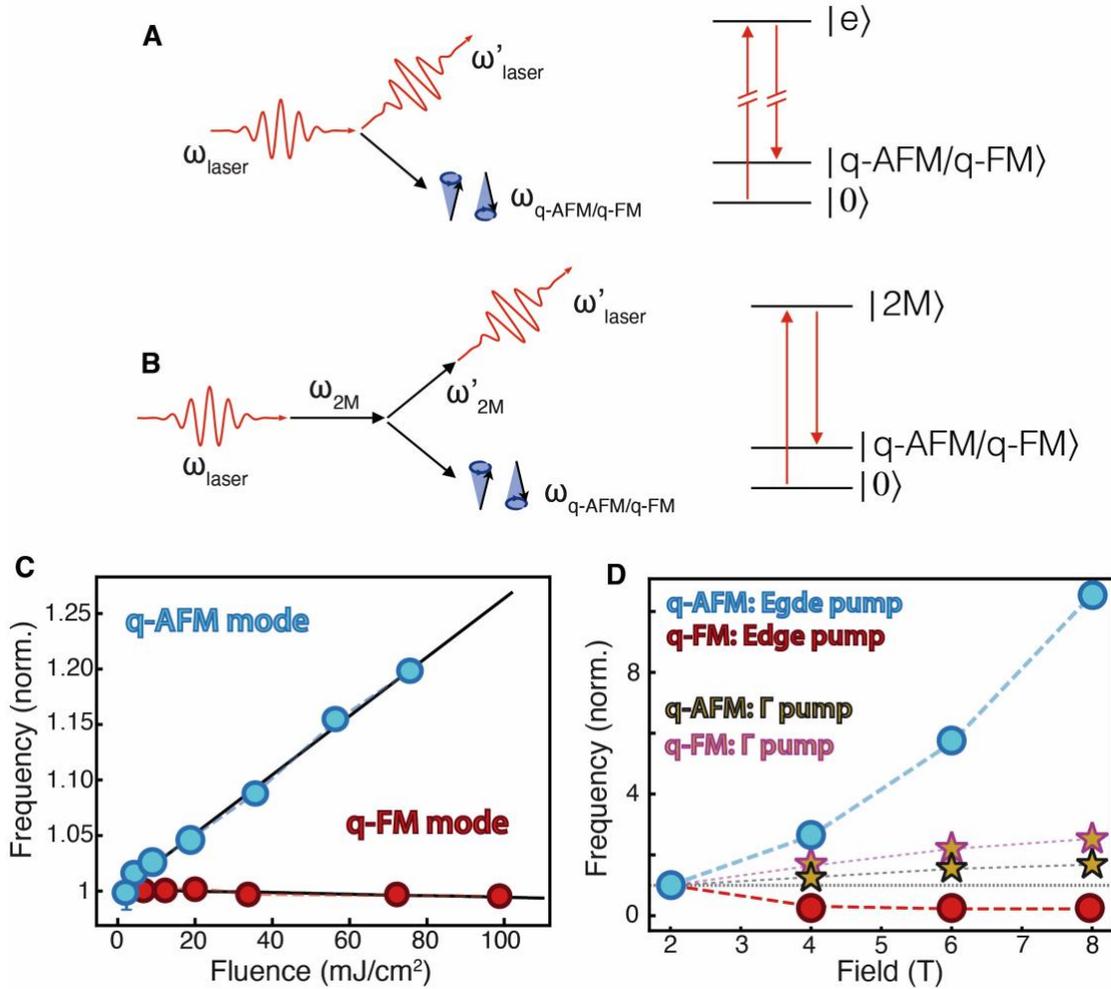

**Fig. 4. Modelling, data and simulations of the spectrum renormalization.** (**A**) In conventional ISRS, a photon gets scattered by the q-FM/q-AFM mode, according to the inverse Faraday or inverse Cotton-Mouton effect (*8*, *31*). The energy diagram of ISRS shows the intermediate state given by a virtual or real electronic state |e⟩. (**B**) In 2MRRS, a photon gets absorbed by the 2M excitation, which then scatters the q-FM/q-AFM magnon and re-emits a photon with a different frequency, leading to an effective Raman scattering mechanism, resonantly enhanced by the 2M mode. In the energy diagram of 2MRRS the intermediate state is given by the 2M excitation, |2M⟩. (**C**) Fluence dependence of the normalized frequencies of the Gamma-point modes. This plot is extracted from the data in Supplementary Fig. S10 and S11. (**D**) Simulated frequency shift of the Gamma-point modes, as a function of the intensity of the fictitious magnetic field, induced exciting either high-momentum ("Edge pump", circles) or Gamma-point ("Γ pump", stars) magnons. The experimental findings are qualitatively reproduced: driving high-momentum magnons induces frequencies shifts with the correct sign (circles). In contrast, a simple anharmonicity of the Gamma-point magnons, without exciting high-momentum modes, leads to a blueshift of both the q-FM and q-AFM modes (stars), in disagreement with our experimental results. The dashed grey line is set to 1, which is the unperturbed value of the normalized eigenfrequency.

# Supplementary Materials for

**Dynamical renormalization of the magnetic excitation spectrum via high-momentum nonlinear magnonics**

C. Schönfeld[1], L. Feuerer[1], J. Bär[1], L. Dörfelt[1], M. Kerstingskötter[1], T. Dannegger[1], D. Wuhrer[1], W.Belzig[1], U. Nowak[1], A. Leitenstorfer[1], D. Juraschek[2,3] and D. Bossini[1]*

[1]Department of Physics and Center for Applied Photonics, University of Konstanz, D-78457 Konstanz, Germany

[2]School of Physics and Astronomy, Tel Aviv University, Tel Aviv 6997801, Israel.

[3]Department of Applied Physics and Science Education, Eindhoven University of Technology, Eindhoven, Netherlands.

Corresponding author: davide.bossini@un-konstanz.de

**The PDF file includes:**

> Materials and Methods
> Supplementary Text
> Figs. S1 to S#
> Tables S1 to S#

## Materials

### Sample and characterization

At room temperature hematite is a canted antiferromagnet, comprising two $Fe^{3+}$ sublattices coupled in an antiparallel fashion by the exchange interaction $\boldsymbol{S}^{\Uparrow}$ and $\boldsymbol{S}^{\Downarrow}$ in Fig. 1**A**. The Dzyaloshinski-Moriya interaction (DMI) cants the spins, inducing a net magnetization in the *ab*-plane ($\boldsymbol{M = 5 \times 10^{-4}}$emu in Fig. 1**C** and Fig. S1). For our experiments we employed a single-crystal, cut along the [001] axis. The crystal was commercially purchased and polished down to a 30-μm thickness, which matches the penetration depth of light in the center of the 2M mode at 45 THz. We measured the absorption spectrum of our sample, as shown in Fig. 1**B** with a commercial FTIR spectrometer. The peak value of the absorption coefficient in at the 2M mode is 125 cm⁻¹ taking into account also the reflectivity of both hematite surfaces. A magnetic characterization was also performed, as the hysteresis loop of the magnetization as a function of an applied magnetic field was measured at 300 K with a superconducting quantum interference device (SQUID).

The data (Fig. S1) demonstrate that the 200 mT field applied during our pump-probe experiments is intense enough to saturate the magnetization and, therefore, to perform an experiment under stroboscopic conditions.

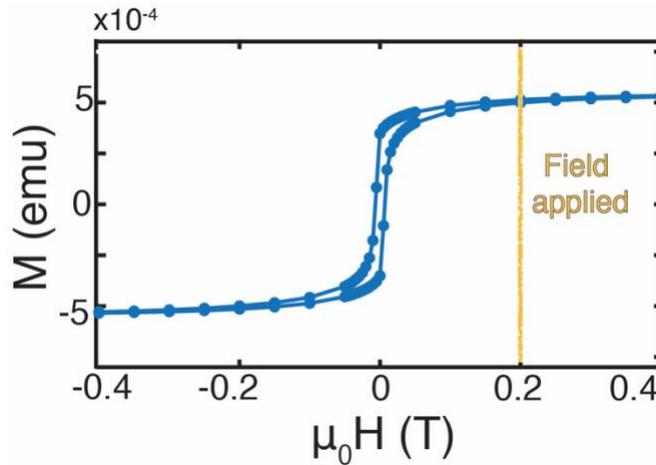

**Fig. S1. Magnetic hysteresis loop of hematite.** The room temperature SQUID measurements show that the hysteresis loop is closed, if a field of 200 mT is applied. This is the value of the static magnetic field employed in our pump-probe experiments.

### Two-magnon absorption in Hematite

The early observations of the 2M mode in the absorption spectrum of magnetic materials demonstrated this process to be of electric-dipole origin (*18*, *37*) Following the original and canonical treatment of the problem (*19–21*), let us introduce the light-matter interaction as

$$H_{2M} = -\mathbf{p}_{eff} \cdot \mathrm{E} \tag{S3}$$

where $\mathbf{p}_{eff}$ is the effective electric dipole moment and $\mathbf{E}$ represents the electric field component of light. Taking the conservation of spin into account, the lowest-order contribution to the effective electric dipole is given by

$$\mathbf{p}_{eff} = \Pi_{ij}\big(\mathrm{S_i} \cdot \mathrm{S_j}\big) \tag{S4}$$

where $i,j$ indicate ionic sites in the crystal, the form of the tensor $\Pi_{ij}$ is obtained from consideration of the crystal symmetry and $\mathbf{S}_{i,j}$ are spins located at the corresponding ionic sites. Considering nearest neighbor spins in multisublattice materials, it follows that $\mathbf{S}_i$ and $\mathbf{S}_j$ belong to two different sublattices. This latter consideration in combination with the form of Eq. (S2) reveals why the microscopy of $\Pi_{ij}$ is related to the exchange interaction (*19–21*). Considering the pair of ions $i$ and $j$ belonging to two different sublattices ($\Uparrow$ and $\Downarrow$) where a spin-up and a spin-down electrons are accommodated in orbitals $\varphi_i$ and $\varphi_j$, respectively. Formally, we can now express the transition electric dipole moment accompanying the flip of both spins as

$$\langle\varphi_{i\Uparrow}\varphi_{j\Downarrow}|P_{eff}|\varphi_{j\Uparrow}\varphi_{i\Downarrow}\rangle = \sum_{\mu}\langle\varphi_{i\Uparrow}\varphi_{j\Downarrow}|\mathbf{P}|\varphi_{\mu\Uparrow}\varphi_{j\Downarrow}\rangle\langle\varphi_{\mu\Uparrow}\varphi_{j\Downarrow}|V|\varphi_{j\Uparrow}\varphi_{i\Downarrow}\rangle/\Delta E(\varphi_\mu\leftarrow\varphi_i)\ + \tag{S5}$$

$$+ \sum_{\nu}\langle\varphi_{i\Uparrow}\varphi_{j\Downarrow}|\mathbf{P}|\varphi_{i\Uparrow}\varphi_{\nu\Downarrow}\rangle\langle\varphi_{i\Uparrow}\varphi_{\nu\Downarrow}|V|\varphi_{j\Uparrow}\varphi_{i\Downarrow}\rangle/\Delta E(\varphi_\nu\leftarrow\varphi_j),$$

where $\mathbf{P}$ is the electric-dipole moment operator, $\Delta E(\varphi_{\mu,\nu}\leftarrow\varphi_{i,j})$ is the energy required for transferring an electron from $\varphi_{i,j}$ to $\varphi_{\mu,\nu}$, where $\varphi_{\mu,\nu}$ are any odd-parity orbitals in the excited state of the ion $i$ and $j$, so that an allowed electric-dipole transition can occur. In the most general case ($\varphi_\mu \neq \varphi_j, \varphi_\nu \neq \varphi_i$), the interaction operator $V$ is the two-electron Coulomb operator $r_{12}^{-1}$. Consequently, the matrix elements of $V$ are the canonical exchange integrals, more specifically

$$\langle\varphi_{\mu\Uparrow}\varphi_{j\Downarrow}|V|\varphi_{j\Uparrow}\varphi_{i\Downarrow}\rangle = \int dr_1\, dr_2\, \varphi_\mu^*(r_1)\varphi_j^*(r_2)V\varphi_j(r_1)\varphi_\mu(\mathbf{r_2}) \tag{S6}$$

Figure S2 graphically displays this process. The states labelled as 1 and 2 represent $\varphi_{i,j}^{\Uparrow,\Downarrow}$: considering the spin orientation determined in the ground state of the system by the exchange interaction, flipping a spin while keeping the electron in the same orbital generates the $\Delta E_{1-2}$ energy split. An analogous consideration applies to the energy split of the states 3 and 4, which describe an orbital excited state. The photon energy required to induce the absorption of the magnon pair is $h\nu = 2\Delta E_{1-2}$. The blue arrow represents the spin-preserving electric dipole transition $\langle\varphi_{i\Uparrow}\varphi_{j\Downarrow}|\mathbf{P}|\varphi_{\mu\Uparrow}\varphi_{j\Downarrow}\rangle$ in Eq. (S3), while the red arrows depict the transitions driven by the potential $V$. Importantly, the spin-preserving transition can be either a real electric dipole transition between two orbitals of the magnetic ions or the charge-transfer transition, depending on the electronic structure of the material under investigation.

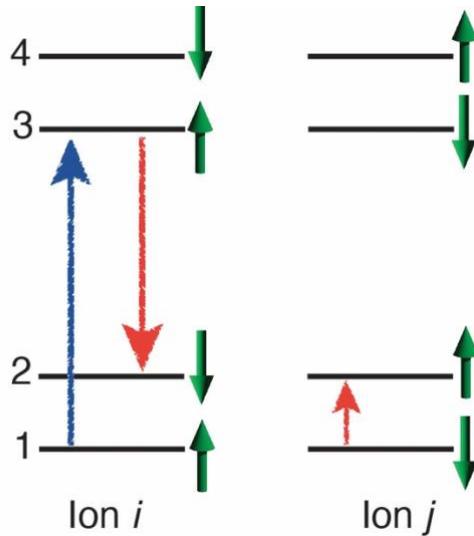

**Fig. S2. Exchange mechanism.** The series of transitions described in the text enabling the generation of magnon pairs, via the absorption of the electric field of light, is displayed.

## Methods

### Experimental set-up for mid-infrared magneto-optical pump-probe experiments

Our self-developed and built table-top laser system combines femtosecond Er:Fiber and Yb:Fiber technology with a high-power regenerative Yb:thin-disk amplifier. The proper thermal management of the amplification geometry(*38*) allows for nearly transform-limited 615 fs pulses with central wavelength of 1030 nm and 17 mJ pulse energy at a repetition rate of 3 kHz (*39*). The output of the laser system drives first a nonlinear frequency conversion consisting of a white-light (WL) seed, generated in a YAG plate, and a subsequent three-stage optical parametric amplifier (OPA). Together with the remaining components of the Yb:thin disk amplifier (pump), the redshifted beam (signal) provided by the OPA generates intense multi-THz pulses (idler) in AgGaS$_2$ via difference-frequency mixing. The idler intensity is conveniently adjustable at 45 THz central frequency by polarization-based intensity adjustment of the pump. The central frequency and bandwidth of the idler beam can be fine-tuned by varying the signal spectrum and the phase matching angle of the 2 mm-thick AgGaS$_2$ crystal. A parallel WL and OPA sequence generate the near-infrared probe beam spectrally covering a 56 THz-broad band at 1/e$^2$ width. The duration of the probe pulses is sub-20 fs. This probe beam propagates collinearly with the mid-infrared beam and both are focused in a 15 μm-thin GaSe crystal for field-resolved detection of the mid-infrared output by electro-optic sampling. Calculations considering the maximum achievable fluence and detected electric-field waveform yield a peak electric field strength exceeding 300 MV/cm. All propagation paths of the mid-infrared beam are held in nitrogen atmosphere to avoid molecular absorption by e.g. water vapor. The electric field and spectrum of our mid-infrared laser pulses are shown in Fig. S3.

The ultrafast spin dynamics is detected by monitoring the changes of the ellipticity of the probe beam in a balanced detection scheme, comprising a quarter-wave plate, a Wollaston prism and two photodiodes. The same detection scheme has been employed to detect coherent low-energy magnons in hematite(*22*). In particular, it was demonstrated that in our configuration the magneto-optical Cotton-Mouton effect generates the detected ellipticity of the probe polarization(*22*).

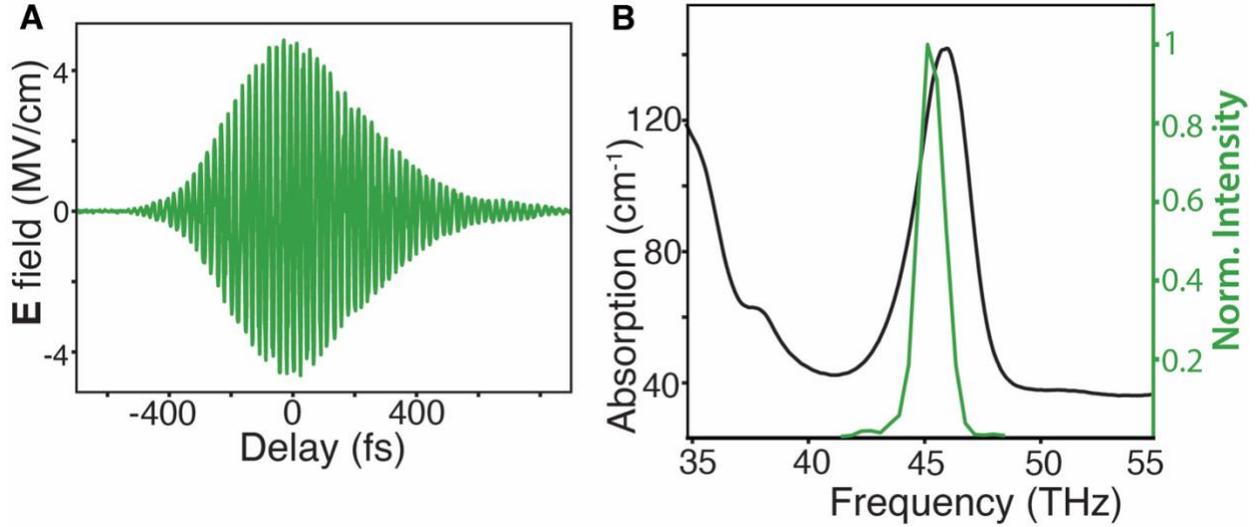

**Fig. S3. Mid-infrared laser pulses.** (**A**) Electric field trade of the femtosecond laser pulses with central frequency of 45 THz, as employed to measure the green graph in Fig. 2. The field corresponds to a fluence of 6 mJ/cm². The field was measured via electro-optical sampling with the probe beam, employed also to detect the transient spin dynamics. (**B**) Green graph: Spectrum corresponding to the electric field transient in panel (**A**), shown on top of the absorption spectrum of hematite (black). Note that the bandwidths of the pump pulses and the 2M mode are similar.

**Data analysis**

The pump-probe data were fitted using a function comprising the sum of two exponentially damped sinusoidal terms, as well as an exponential decay term, namely

$$f(t) = \sum_{i=1}^{2} a_i \sin(2\pi \nu_i t + \phi_i) \exp\left(-\frac{t}{\tau_i}\right) + b_{\text{bk}} \exp\left(-\frac{t}{\tau_{bk}}\right), \qquad \text{(S7)}$$

where $a_i$, $\nu_i$, $\phi_i$ and $\tau_i$ refer to the amplitude, frequency, phase and relaxation time of the q-FM and q-AFM modes, respectively. The last term describes the background on top of which the oscillations are superimposed. We determine $\tau_{bk}$ to be on the order of 100-150 ps throughout all datasets shown in the manuscript and Supplementary Materials. The overall fit quality is excellent, as demonstrated by R-squared values ranging between 0.95 and 0.99. Furthermore, error bars are also extracted from the fitting procedure and represent the confidence bounds (95%) of the coefficients.

## Supplementary Note 1: Theoretical basis for the excitation mechanism

We propose a mechanism based on the resonant enhancement of the efficiency of Raman scattering by the quasi-antiferromagnetic (q-AFM) and quasi-ferromagnetic (q-FM) modes, mediated by the two-magnon excitation. In the following, we will recall the theoretical formalism for impulsive stimulated Raman scattering and then develop in analogy a formalism for the resonant enhancement by the 2M mode, which we call two-magnon resonant Raman scattering (2MRRS)

**Conventional impulsive stimulated Raman scattering (ISRS)**

The fundamental Raman-type light-matter interaction mechanism of magnon modes with ultrashort laser pulses is impulsive stimulated Raman scattering (ISRS) (*31*, *40*, *41*). The interaction Hamiltonian is given by the coupling of the electric dipole moment, **p**, with the electric field component of light, **E**,

$$H_{int} = -\mathbf{p} \cdot \mathbf{E}. \tag{S8}$$

The first-order electric dipole moment for Raman scattering is given by

$$p_i = \frac{\varepsilon_0}{V} \chi_{ij} E_j \tag{S9}$$

Here, $\varepsilon_0$ is the vacuum permeability, $V$ is the volume of the unit cell and $\chi_{ij}$ is the linear electric susceptibility. We use the Einstein sum convention in the following. Phenomenologically, the electric dipole moment produced by magnon modes can be written as an expansion of the electric susceptibility in terms of the components of the magnetization and antiferromagnetic vectors, **M** and **L**. For an antiferromagnet with two sublattices, where $M = S^{\Uparrow} + S^{\Downarrow}$ and $L = S^{\Uparrow} - S^{\Downarrow}$, the expansion takes the general form (*42*, *43*)

$$\chi_{ij} = \chi_{ij}^0 + \alpha_{ijk}M_k + \beta_{ijkl}M_kM_l + \beta'_{ijkl}M_kL_l + \beta''_{ijkl}L_kL_l + \cdots \tag{S10}$$

The mechanism corresponding to terms linear in $\mathbf{M_i}$ is called the inverse Faraday effect, whereas that corresponding to quadratic terms is called the inverse Cotton-Mouton effect (*31*, *44*). A typical scattering and energy diagram can be found in Fig. 4**A** and **B** of the main text. The coherent spin dynamics of the Gamma-point magnons can be captured by the sublattice-resolved form of the Landau-Lifshitz-Gilbert equations (*45*, *46*),

$$\frac{d\mathbf{S_s}}{dt} = -\frac{\gamma_{el}}{1 + \kappa_{el}^2} \left[ \mathbf{S_s} \times \mathbf{B_s}^{eff} - \frac{\kappa_{el}}{|\mathbf{S_s}|} \mathbf{S_s} \times \left( \mathbf{S_s} \times \mathbf{B_s}^{eff} \right) \right] \tag{S11}$$

where $\mathbf{B_s}^{eff}$ is the effective magnetic field acting on the sublattice spins, given by $\mathbf{B_s}^{eff} = -\partial H/(\partial \mathbf{S_s})$, where $H = H_0 + H_{int}$ is the total spin Hamiltonian, containing both the ground-state Hamiltonian, $\mathbf{H_0}$, and the interaction Hamiltonian, $\mathbf{H_{int}}$. The electronic gyromagnetic ratio is indicated by $\mathbf{\gamma_{el}}$, $\mathbf{\kappa_{el}}$ is the phenomenological Gilbert damping, and the length of the spins is normalized to $|\mathbf{S_s}| = \mathbf{1}$.

**Two-magnon resonant Raman scattering (2MRRS)**

We now describe a mechanism, by which the Raman scattering efficiency is enhanced through the two-magnon resonance. A photon is absorbed by the 2M excitation, which then scatters by the q-AFM or q-FM mode and re-emits a photon with modified frequency. The scattering and energy diagrams of the process are shown in Figs. 4A and B.

For 2M excitations, the macrospin approximation for the magnetization and antiferromagnetic vectors is not sufficient however (*17*, *30*, *32*).The excitation is instead described by a spin-correlation function, $C_{ij,k} \sim \langle S_{i,k} S_{j,-k} \rangle$, where $k = \pi/a$ in our case of excitation of high-momentum magnons in hematite, where **a** represents the lattice constant. Because the electric dipole moment of the 2M excitation is given by $p_{i,k} = \Pi_{ijk,k} S_{j,k} S_{k,-k}$ (*20*), we can write the interaction potential for the absorption process as

$$H_{int} = \Pi_{ijk,\pm k} C_{ij,\pm k} E_k \qquad (S12)$$

The resonantly driven 2M excitation then couples nonlinearly to the q-AFM and q-FM magnons, which we again can express in terms of the macrospin approximation,

$$H_{nonlinear} = \kappa_{ijk,\pm k} C_{ij,\pm k}^2 M_k + \kappa'_{ijkl,\pm k} C_{ij,\pm k}^2 L_k L_l \qquad (S13)$$

Equation (S11) follows the phenomenology of impulsive stimulated Raman scattering in Eq. (S8). The macroscopic mechanisms can therefore be considered inverse Faraday (*8*) and inverse Cotton-Mouton (*31*) effects resonantly enhanced by the 2M excitation, consistent with the experimental findings. The vector of the spin correlation, $C$, oscillates with a frequency of the 2M excitation, $C(t) \sim \sin(\omega_{2M} t)$, where $\omega_{2M} = 2\omega_{\pm \pi/a}$. That means the square of $C$ oscillates with sum- and difference-frequency components that are provided by its own linewidth, $C^2(t) \sim \sin((\omega_{2M} - \omega'_{2M})t) + \sin((\omega_{2M} + \omega'_{2M})t)$. The frequencies of the q-AFM and q-FM magnons at the $\Gamma$ point fall within the linewidth of the difference-frequency component generated by the square of the 2M excitation. This mechanism effectively describes a Raman scattering process, where the Raman scattering efficiency is determined by two factors: 1) the dipole-interaction tensor of the 2M excitation, $\Pi$, and 2) the nonlinear 2M-2M-q-AFM/q-FM coupling tensors, $\kappa$ and $\kappa'$. Note that our experimental methodology does not allow to retrieve the numerical values of these coupling tensors. The proposed mechanism of 2MRRS is further consistent with a comparison of the timescales. Impulsive excitation can occur provided that the stimulus is shorter than the period of the mode to be excited. This condition applies to our case, since the dephasing time of the 2M mode ($\approx 400$ fs) is shorter than the oscillation period of both the q-AFM ($\approx 6$ ps) and q-FM ($\approx 50$ ps) modes. This nonlinear coupling leads to a 2M-mediated contribution to the third-order nonlinear electric susceptibility $\boldsymbol{\chi^{(3)}}$. We show a schematic of the excitation pathway in Fig. S4.

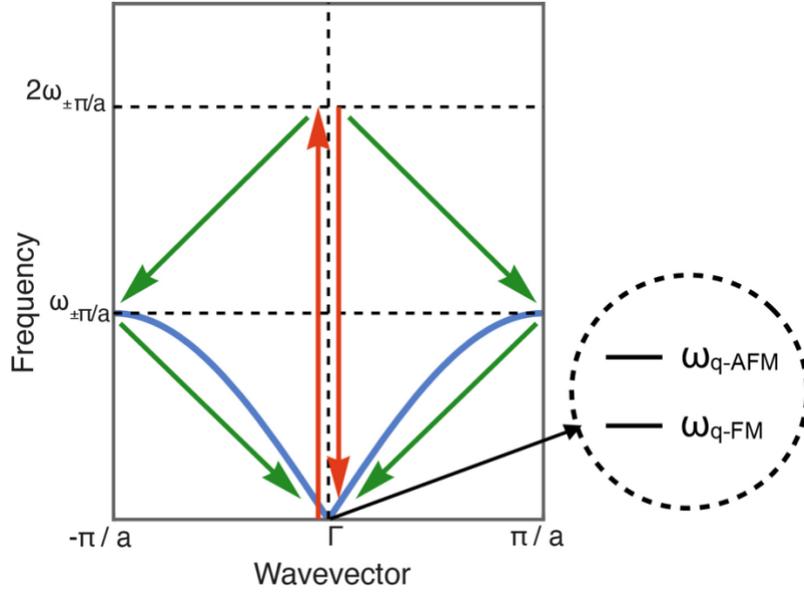

**Fig. S4. Excitation mechanism**. Schematic of the spin wave dispersion (blue curve) and two-magnon resonant Raman scattering (2MRRS). Ultrashort mid-IR pulses resonantly excite pairs of magnons near opposite edges of the Brillouin zone with frequencies $\omega_{\pm \pi/a}$ via mid-IR absorption (red up-arrow and top green arrows). Afterwards, the 2M excitation couples to the q-AFM and q-FM modes with frequencies $\omega_{q-AFM}$ and $\omega_{q-FM}$ at the center of the Brillouin zone, under mid-IR emission (bottom green arrows and red down-arrow).

The dynamics of the 2M spin-correlation function is given by the Heisenberg equation of motion (*30*)

$$\hbar \frac{\partial \mathbf{C}}{\partial t} + \mathbf{C} \times \frac{\partial H}{\partial \mathbf{C}} = 0 \tag{S14}$$

whereas the macrospin dynamics of the q-AFM and q-FM magnons are again described by the Landau-Lifshitz-Gilbert equations, Eq (S12). The total spin Hamiltonian is expressed as $\boldsymbol{H} = \boldsymbol{H_0} + \boldsymbol{H_{int}} + \boldsymbol{H_{nonlinear}}$, and the spin-correlation function drives the Landau-Lifshitz-Gilbert equations as an effective magnetic field $\boldsymbol{B}^{\text{eff}}_s = -\partial \boldsymbol{H}_{nonlinear}/(\partial \boldsymbol{S_s})$. This type of coupling process is different from the three-magnon coupling mechanism observed recently(*9, 14*) which is based on magnetic dipole coupling. We further note that a coherent excitation of the 2M-state has recently been achieved through Raman scattering instead of absorption of light (*17, 30*). This would make a possible nonlinear coupling between the high-momentum modes and the Gamma-point magnons in Eq. (S11) an effective four-photon process however, which we expect to be much weaker than the Raman scattering process described here.

The 2M-enhanced inverse Faraday and inverse Cotton-Mouton effects can be seen in the same light as the phonon inverse Faraday and phonon inverse Cotton-Mouton effects described recently (*44*, *47*, *48*), where optical phonons, instead of a 2M excitation act as a resonant transducer.

## Supplementary Note 2: Analytical theory for spectrum renormalization.

We sketch here the principal aspects underlying our analytical approach to the renormalization of the q-FM and q-AFM frequencies.

We consider a model system which describes a magnet with two sublattices characterized by the Hamiltonian

$$\hat{H} = \sum_{i,j} J_{ij} \hat{\vec{S}}_i \cdot \hat{\vec{S}}_j + K \sum_i \widehat{S_{i,z}^2} - B_{DM} \sum_i \widehat{S_{i,x}} \tag{S15}$$

The Heisenberg exchange is antiferromagnetic and only applies to nearest neighbors ($J_{ij} = J > 0$). The anisotropy is chosen such that it's larger than zero ($K > 0$) and the spins prefer an alignment orthogonal to the anisotropy direction. Thus, the spins form an antiferromagnetic configuration in the system's $x - y$ -plane. $\hat{\vec{S}}_i$ is the spin operator that describes the spin at the lattice site $\vec{r_i}$ with a spin length of $S$. We assume $\hbar = 1$. The last term represents an effective description of the DMI interaction, in terms of an internal field canting the two antiferromagnetically coupled sublattices.

In the absence of an internal field ($B_{DM} = 0$) any collinear antiferromagnetic configuration of spins in the $ab$ plane ($x - y$ coordinates) is equally favourable. If the field is smaller than the critical field strength ($0 < B_{DM} \leq B_{DMcrit} = 2SJ_0$, with $\sum_i J_{ij} = \sum_j J_{ij} = J_0$), the spins tilt away from the $y$-axis along the $x$-axis, forming a field-polarised state along the $x$-axis when the field strength is larger or equal to $B_{DMcrit}$. Our focus will be on $0 < B_{DM} \leq B_{DMcrit}$. The spin operators can be expressed in terms of bosonic creation and annihilation operators in the Holstein-Primakoff transformation

$$\widehat{S_{i,+}} = \sqrt{2S} f(\widehat{n_i^A}) \widehat{a_i}; \ \widehat{S_{i,-}} = \sqrt{2S} \widehat{a_i^\dagger} f(\widehat{n_i^A}); \ S_{i,3} = S - \widehat{a_i^\dagger} \widehat{a_i}; \ i \in A$$

$$\widehat{S_{i,+}} = \sqrt{2S} f(\widehat{n_i^B}) \widehat{b_i}; \ \widehat{S_{i,-}} = \sqrt{2S} \widehat{b_i^\dagger} f(\widehat{n_i^B}); \ S_{i,3} = S - \widehat{b_i^\dagger} \widehat{b_i}; \ i \in B \tag{S16}$$

where the unit vectors $\boldsymbol{e_{3,\sigma}}$, with $\boldsymbol{\sigma} \in \{A, B\}$, are chosen as quantisation directions of the spins in the corresponding sublattices ($A, B$). The number operators are defined in the usual $\widehat{n_i^A} = \widehat{a_i^\dagger} \widehat{a_i}$, $\widehat{n_i^B} = \widehat{b_i^\dagger} \widehat{b_i}$, and $\boldsymbol{f(\widehat{n})} = \sqrt{1 - \widehat{n}/(2S)}$. The creation ($\widehat{a_i^\dagger}$, $\widehat{b_i^\dagger}$) and annihilation operators ($\widehat{a_i}$, $\widehat{b_i}$) fulfil the bosonic commutation relations.

$$\left[ \widehat{a_i}, \widehat{a_j^\dagger} \right] = \left[ \widehat{b_i}, \widehat{b_j^\dagger} \right] = \delta_{ij}; \left[ \widehat{a_i}, \widehat{a_j} \right] = \left[ \widehat{b_i}, \widehat{b_j} \right] = \left[ \widehat{a_i}, \widehat{b_j^\dagger} \right] = \left[ \widehat{a_i}, \widehat{b_j} \right] = 0 \tag{S17}$$

The spin ladder operators are defined as $\widehat{S_{i,\pm}} = \widehat{S_{i,1}} \pm i \widehat{S_{i,2}}$ with $\widehat{S_{i,\alpha}} = \boldsymbol{e_{\alpha,i}} \cdot \hat{\vec{S}}_i$ and $\alpha \in \{1,2,3\}$. The unit vector $\boldsymbol{e_{\alpha,i}}$ corresponds to $\boldsymbol{e_{\alpha,A/B}}$ depending on whether $i \in A$ or $i \in B$.

The function $f$ in terms of the number operator $\hat{n}$ is interpreted in terms of its Taylor series

$$f(\hat{n}) = 1 - \frac{\hat{n}}{4S} - \frac{\widehat{n^2}}{32S^2} - \cdots \qquad (S18)$$

Depending on the number of magnons $\hat{n}$ compared to the spin length $S$ the series can be terminated for lower or higher orders of $\hat{n}$. The lowest order $f(\hat{n}) \approx 1$ results in the linear spin wave theory. If the occupation number increases, like in the case of thermally or optically excited magnons, we get higher order corrections. We are especially interested in the first order correction

$$f_1(\hat{n}) = -\frac{\hat{n}}{4S} \qquad (S19)$$

With this starting point, we investigate the effect of this correction by considering only terms with four operators with the same momentum. This simplification is necessary to treat the theory analytically. It also implies that only the population of zone-edge magnons is treated, while magnon-magnon scattering processes, which are described by magnon operators with different momenta, are not considered. The corrections to the eigenfrequencies of the q-FM and q-AFM modes result

$$\Delta\varepsilon_{0,\text{q-FM}}(n) - \Delta\varepsilon_{0,\text{q-FM}}(0) \approx -\frac{n_{\frac{\pi}{2}}}{2SN\varepsilon_{0,\text{q-FM}}}\left(|B_{\text{DM}}|^2 - \frac{1}{4}S^2KJ_0\right) \qquad (S20)$$

$$\Delta\varepsilon_{0,\text{q-AFM}}(n) - \Delta\varepsilon_{0,\text{q-AFM}}(0) \approx \frac{n_{\frac{\pi}{2}}}{2SN\varepsilon_{0,\text{q-AFM}}}\left(|B_{DM}|^2 - \frac{3}{4}S^2KJ_0\right)$$

Where the approximation of weak perturbation of the collinear antiferromagnetic order, which well describes the ground state of hematite, was employed. Therefore, the field $B_{DM}$ is smaller than $S\sqrt{KJ_0}$. Taking this aspect into account, the theory predicts a blueshift(redshift) for the q-FM(q-AFM) mode, linearly dependent on the photoinduced magnon population at the edges of the Brillouin zone ($n_{\frac{\pi}{2}}$). This result is not consistent with the observations reported in Fig. 2**C** and 2**D** in the main text. We thus conclude that our experiment cannot be reproduced by considering the effects of the photoinduced magnon population, while preserving all the other assumptions on which linear spin wave theory is formulated.

## Supplementary Note 3: atomistic spin dynamics simulations.

The spin model for the atomistic spin simulations is based on a semi-classical Heisenberg model, which is introduced in the literature (*33*). The Heisenberg Hamiltonian

$$\mathcal{H}_{\text{ex}} = -\frac{1}{2}\sum_{i \neq j} \boldsymbol{S}_i^{\text{T}} J_{ij} \boldsymbol{S}_j \qquad (S21)$$

with the tensors $J_{ij}$, includes the isotropic exchange, the two-ion anisotropy, the dipole-dipole interaction, and the DMI (Dzyaloshinskii-Moriya interaction), which causes hematite to be a weak ferromagnet above the Morin transition.

The Heisenberg Hamiltonian is extended by the terms of second- and fourth-order anisotropy with

$$H_{oa} = -d_2 \sum_i S_{i,z}^2 - d_4 \sum_i S_{i,z}^4 \tag{S22}$$

To excite magnons with an external magnetic field $B$, the Zeeman term

$$\mathcal{H}_{\mathrm{Zee}} = -\mu_s \sum_i \boldsymbol{B}_i \cdot \boldsymbol{S}_i \tag{S23}$$

is added to the Hamiltonian. All parameters used are described in (*33*). The dynamics of the individual spins are described by the stochastic Landau-Lifshitz-Gilbert equation (*49–51*)

$$\frac{\mathrm{d}}{\mathrm{dt}} \mathbf{S}_i = -\frac{\gamma}{(1+\alpha^2)\mu_s} \mathbf{S}_i \times (\mathbf{H}_i + \alpha \mathbf{S}_i \times \mathbf{H}_i) \tag{S24}$$

with $\mathbf{H}_i(t) = \boldsymbol{\zeta}_i(t) - \frac{\partial \mathcal{H}}{\partial \mathbf{S}_i}$. Here $\boldsymbol{\zeta}_i(t)$ is a stochastic field, which describes the coupling of the system to a heat bath.

For our simulations we choose the damping parameter $\alpha = 0.001$. This value is larger than experimentally determined value (*52*), but it leads to a faster relaxation towards equilibrium. This difference also reduces the resonance frequencies slightly but only by 3 %. To numerically obtain the dispersion relation, we use a Fourier transform of a complex-valued combination of the cartesian components of the normalized sublattice magnetization vector of one of the four sublattices, averaged over planes perpendicular to the *k*-vector. This signal is constructed differently below and above the Morin temperature $T_M$ and is given by

$$\langle S_x \rangle + i \langle S_y \rangle \text{ if } T < T_M$$
$$\langle S_z \rangle + i \langle S_y \rangle \text{ if } T > T_M$$

(S25)

The numerically calculated dispersion relation is shown in Fig. S5. The branches in the middle, which show the typical $f \propto |\sin(k)|$ dependence, are the q-AFM and q-FM modes, although their splitting is not visible in the plot. The upper modes can be identified as the exchange modes. The difference in intensity for positive and negative frequencies reflects the chirality of the modes.

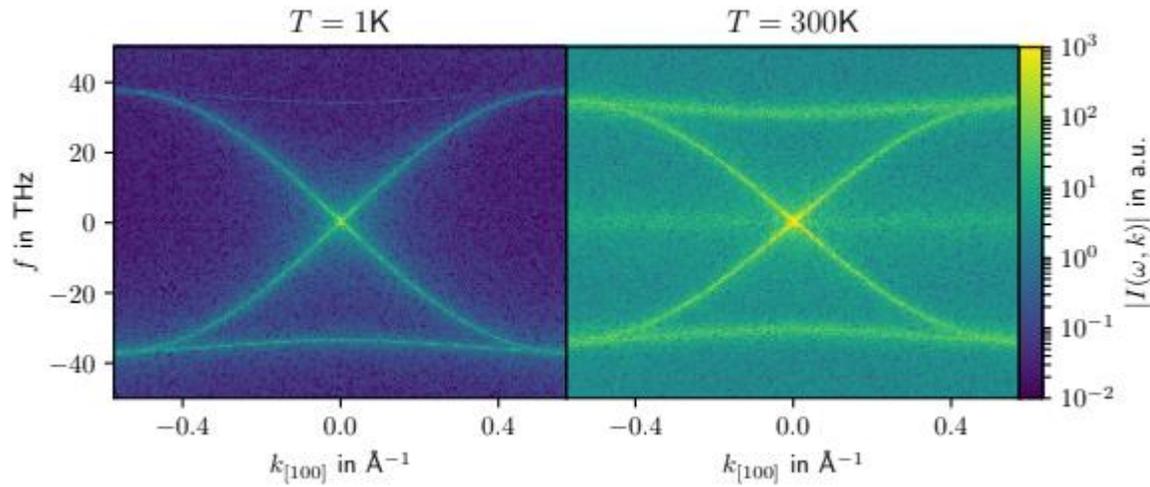

**Fig. S5. Spin wave dispersion relation hematite**. Thermal excitation of the spin waves at $T = 1$ K and $T = 300$ K.

## Simulations exciting zone-edge magnons

To simulate the effect of spin waves excited at the edges of the Brillouin zone, we rely on the fictitious magnetic field

$$\boldsymbol{B}_i = \left(0, B_{0,y} \cos(\omega t - \boldsymbol{k}\,\boldsymbol{r}), B_{0,z} \cos(\omega t - \boldsymbol{k}\,\boldsymbol{r} - 90°)\right)^T \delta_{iA}, \qquad \text{(S26)}$$

which is zero on sublattices B, C, D and oscillates on sublattice A with a $k$ vector corresponding to the edge of the Brillouin zone in a particular direction. By applying this oscillating field only on one sublattice, the movement of the other sublattices remains unconstrained and since the frequencies of all four modes lie close together at the zone edge, we expect all of them to be excited simultaneously.

The amplitude $B_{0,y/z}$ is varied in different simulations. We set the temperature to $T = 300$ K $> T_M$, consistent with the experiment. The results are shown in Fig. S6. The upper branch contains the q-AFM mode (at $\boldsymbol{k} = 0$), while the lower branch contains the q-FM mode (at $\boldsymbol{k} = 0$). It is visible from Fig. S6 that the frequency of the q-AFM mode increases as the magnetic field becomes stronger, while the q-FM mode decreases in magnitude. These trends are consistent with the experiments reported in Fig. 2 of the main text.

We fit the modes at $\boldsymbol{k} = 0$ with Lorentzian peaks, to read out the frequency and bandwidth of the magnetic resonance modes. The results of the fitting procedure are the data points shown in Fig. 4**D** in the main text.

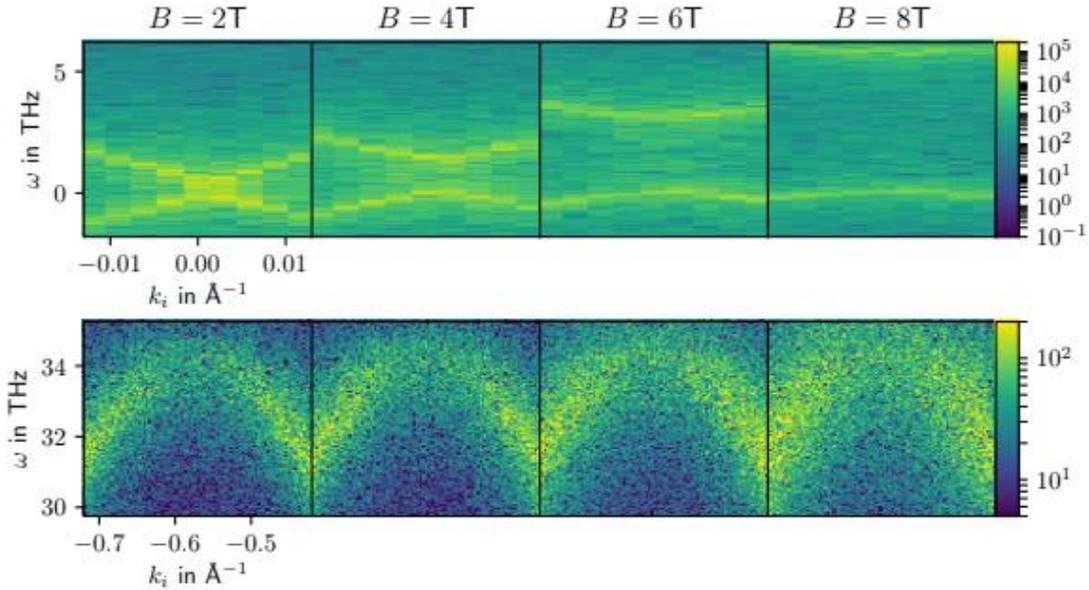

**Fig. S6. Simulated renormalization of the magnon eigenfrequencies driving zone-edge modes.** Monochromatic excitation at T = 300 K. For T > $\boldsymbol{T}_M$. The frequencies of the upper branch increase as a function of the magnetic field, while the frequencies of the lower branch approach zero, implying that the modulus is reduced. The modes at $\vec{\boldsymbol{k}} = \boldsymbol{0}$ of the upper and lower branches correspond to the q-AFM and q-FM modes, respectively.

## Simulations exciting zone-center magnons

We also preformed simulations exciting the q-AFM mode directly at $k = 0$. For this, the wave vector in equation (S26) is set to zero, and the excitation frequency is tuned to match the eigenfrequency of the q-AFM mode. The effect on the band gap is shown in Fig. S7. The frequency of the q-AFM and q-FM modes is determined using Lorentz fits. The results are shown in the main text in Fig. 4**D**. It can be observed that the frequency of the q-AFM mode increases, though much less than by exciting high-momentum spin waves. The frequency of the q-FM mode, on the other hand, increases in magnitude. This behavior also differs from the excitation of the zone edge modes.

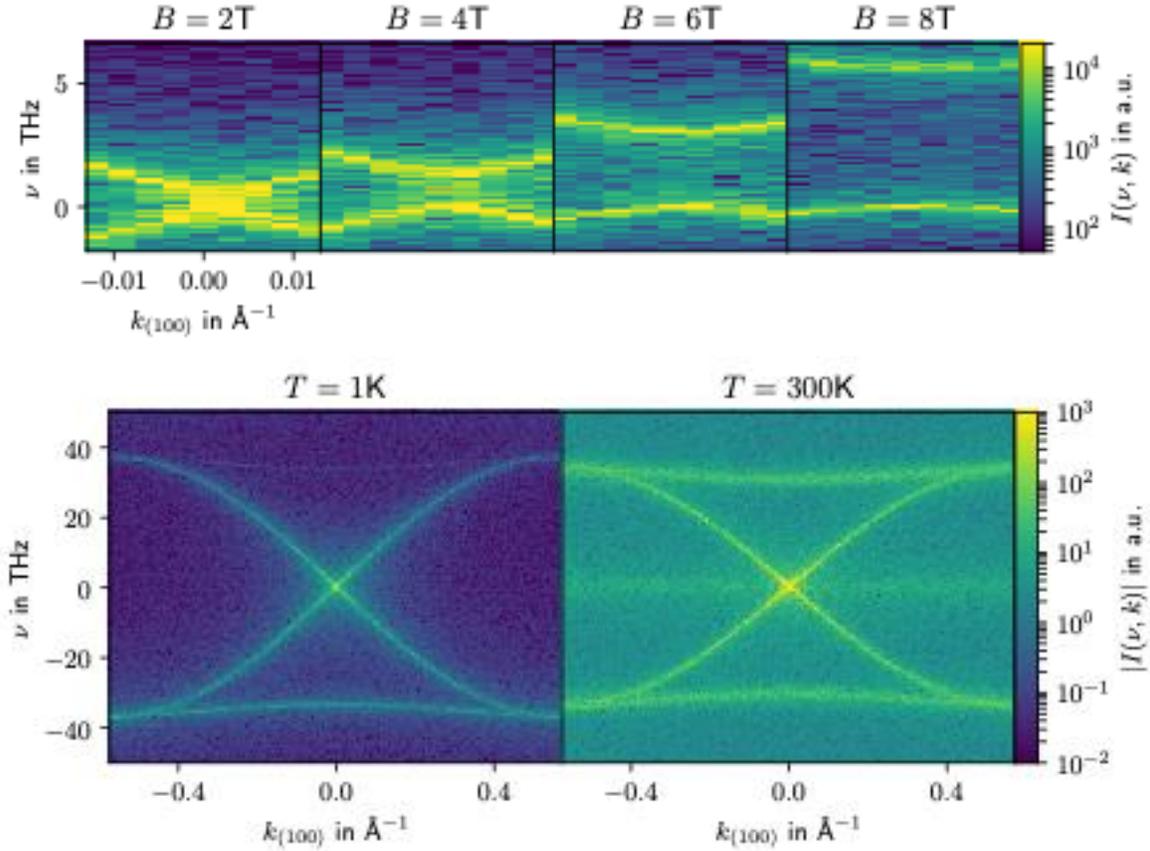

**Fig. S7. Simulated renormalization of the magnon eigenfrequencies driving the q-AFM mode.** Monochromatic excitation at T = 300 K. the frequency of the q-AFM mode increases, though much less than by exciting high-momentum spin waves. The frequency of the q-FM mode, on the other hand, increases in magnitude. This behavior also differs from the excitation of the zone edge modes and from the experimental observations.

## Supplementary Note 4: additional experimental data

**Magnetic field dependence**

Supplementary Figure S8 shows the dependence of the frequencies of the q-AFM and q-FM modes on the magnetic field externally applied. The frequency of the q-AFM is not affected by the magnetic field, as its intensity (< 200 mT) is negligible in comparison with the exchange interaction. On the other hand, the q-FM frequency scales with the field, according to the equation shown in the figure, which expresses the standard magnetic-field dependence of the frequency of the q-FM mode in weak ferromagnets. Using this equation to fit the data, we obtain a value of the gyromagnetic ratio $\gamma = 28.5 \pm 3.5$ GHz/T, which is quantitatively consistent with the literature [26]

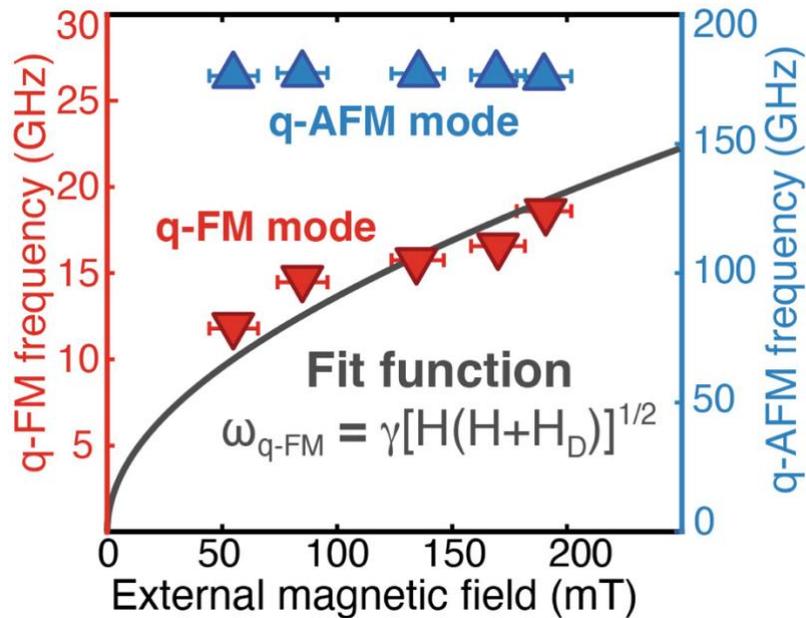

**Fig. S8. Dependence of the frequencies of the q-AFM and q-FM modes on an externally applied magnetic field.** The measurements were performed under the same experimental conditions of the green curve in Fig. 2**A** of the main text. The field dependence of the q-FM mode was fitted with the equation shown in the figure.

**Pump spectral dependence**

The plots shown in Fig. 2**C** and 2**D** represent our core experimental finding. They were obtained by measuring the spin dynamics as a function of the pump frequency, which was systematically scanned from the red- to the blue-side of the 2M mode (centered at $\nu = 45$ THz). The time traces are shown in Fig. S9.

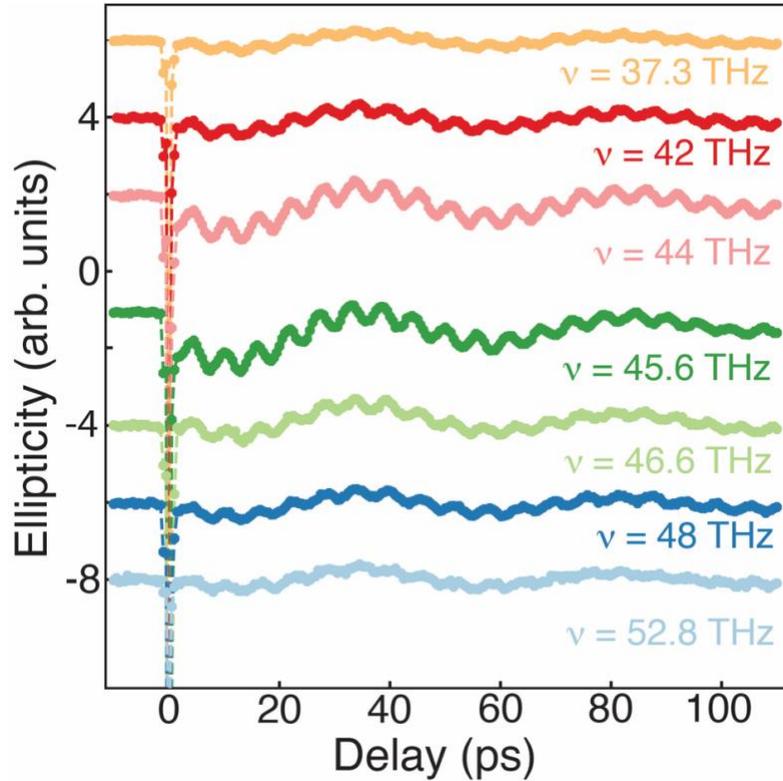

**Fig. S9. Spectral dependence of the spin dynamics.** The real-time spin dynamics as a function of the pump central frequency is shown. The pump central frequency is indicated for each data set. Pump and probe beams were linearly polarized along the direction parallel and orthogonal to the magnetic field, respectively. The fluence was set to 6 mJ/cm$^2$. The transient ellipticity of the probe polarization was monitored in the balanced-detection scheme.

**Fluence dependence**

The dependences of the amplitude and frequency of the q-AFM mode on the fluence for different pump frequencies ("ON" = 45 THz, "OFF" = 39 THz) are reported in Fig. S10**A** and S10**B**, respectively. The amplitude of the q-AFM mode scales linearly (the black lines are fit to the data) with the fluence. The fluence dependence of the frequency is also linear. The modification of the frequency observed via off-resonant pumping is negligible in comparison with resonant pumping. Comparing quantitatively the slopes of the fitting curves (black lines), in the on-resonant case this value is more than 10 times bigger than in the off-resonant case. The minor modification of the frequency observed pumping far away from the resonance is ascribed to single-mode anharmonicity effects of the spin dynamics, due to the large amplitude of the spin oscillations. This effect is negligible in magnitude in comparison with the renormalization of the frequency observed pumping the 2M mode resonantly.

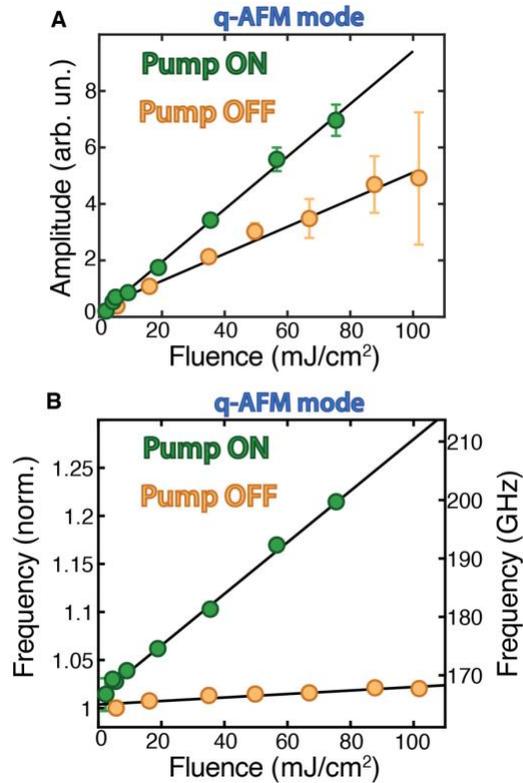

**Fig. S10. Fluence dependence of the q-AFM mode.** Pump and probe beams were linearly polarized along the direction parallel and orthogonal to the magnetic field, respectively.

The dependences of the amplitude and frequency of the q-FM mode on the fluence for different pump frequencies ("ON" = 45 THz, "OFF" = 39 THz) are reported in Fig. S11**A** and S11**B**, respectively. The amplitude of the q-FM mode scales linearly (the black lines are fit to the data) with the fluence. The fluence dependence of the frequency is also linear for the Pump "ON" case, while a clear trend outside the experimental uncertainty cannot be identified in the Pump "OFF" condition.

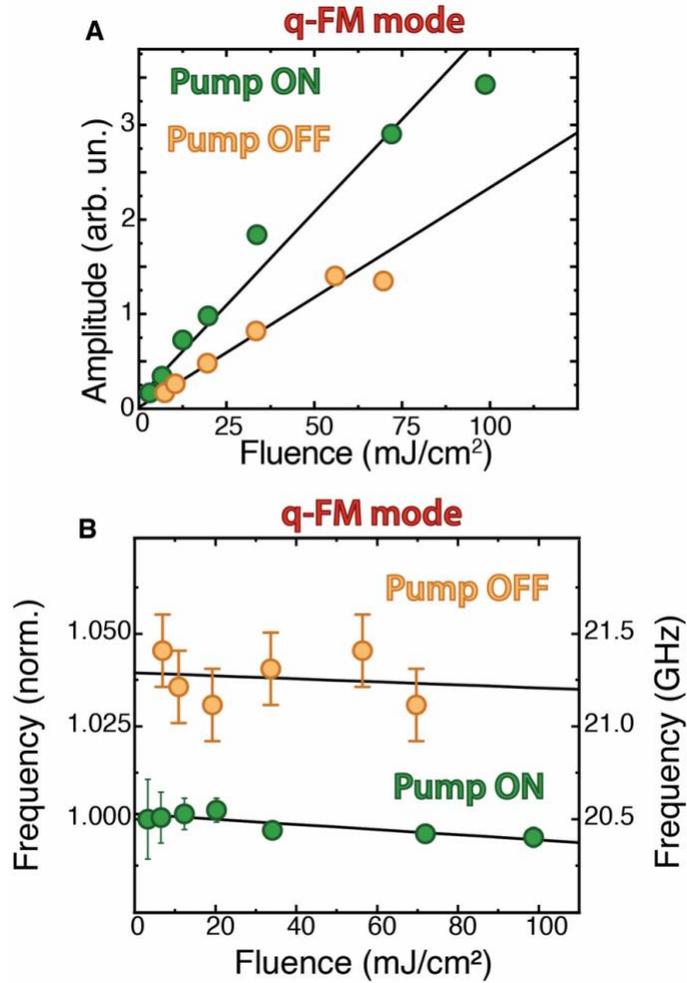

**Fig. S11. Fluence dependence of the q-FM mode.** Pump and probe beams were linearly polarized along the direction parallel and orthogonal to the magnetic field, respectively.

## Pump polarization dependence

The entire dataset describing the dependence of the spin dynamics on the pump polarization is shown in Fig. S12. In order to extract the dependence of the coherent oscillations on the pump polarization, the time-traces in Fig. S12 were fitted and the amplitude of the q-AFM and q-FM modes were plotted in Fig. S13**A** and S13**B**. A periodic dependence of the amplitude of the magnonic oscillations on the angle of the pump polarization with a period of 180° is observed. The periodicity is not affected by the pump frequency, i.e. whether hematite is pumped by driving resonantly or off-resonantly the 2M mode. Therefore, the choice of the pump frequency does not modify the symmetry of the light-matter interaction, implying that in both regimes the same microscopic mechanism (i.e. light scattering) is responsible for the observation.

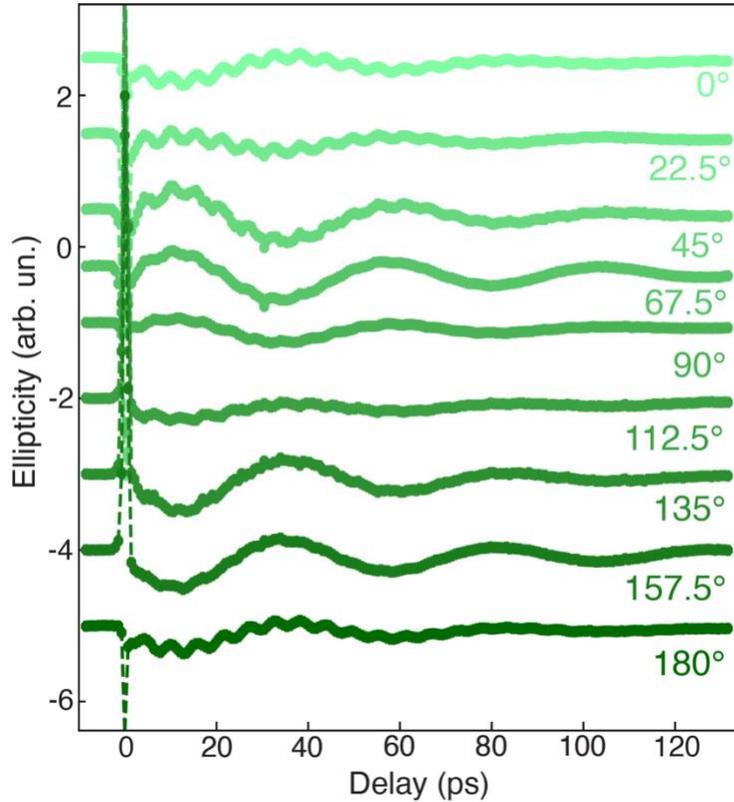

**Fig. S12. Pump polarization dependence of the spin dynamics.** The measurements were performed under the condition of resonant excitation of the 2M mode ("ON" in Fig. S12). The probe beam is linearly polarized along the direction orthogonal to the magnetic field. The polarization of pump beam parallel to the external magnetic field for an angle of 0°. The fluence was set to 6 mJ/cm². The transient ellipticity of the probe polarization was monitored in the balanced-detection scheme.

Additionally, we measured the dependence of the absorption of hematite on the polarization of the pump beam. The absorption of light by hematite is found to be isotropic, both if the pump frequency is resonant and off-resonant with the 2M mode (see Fig. S13**C**). As the absorption of light is isotropic in the plane of the sample, while the spin dynamics is not, it follows that the excitation mechanism of the coherent magnons cannot be ascribed to dissipative processes. The final and ultimate confirmation of this statement is provided by the measurements performed with different duty cycles (Fig. 3 of the main text).

We have also measured the dependence of the spin dynamics on the helicity of circularly polarized mid-infrared laser pulses (Supplementary Fig. S14). The pump central frequency was set to 45 THz, resonant to the 2M mode. Changing the helicity of the pump beam reverses the phase of the q-FM mode (as highlighted by the magenta dashed line), while leaving the phase of the q-AFM unaffected. We note that this observation is consistent with the phenomenology of the inverse Faraday effect (q-FM mode) and inverse Cotton-Mouton effect (q-AFM mode).

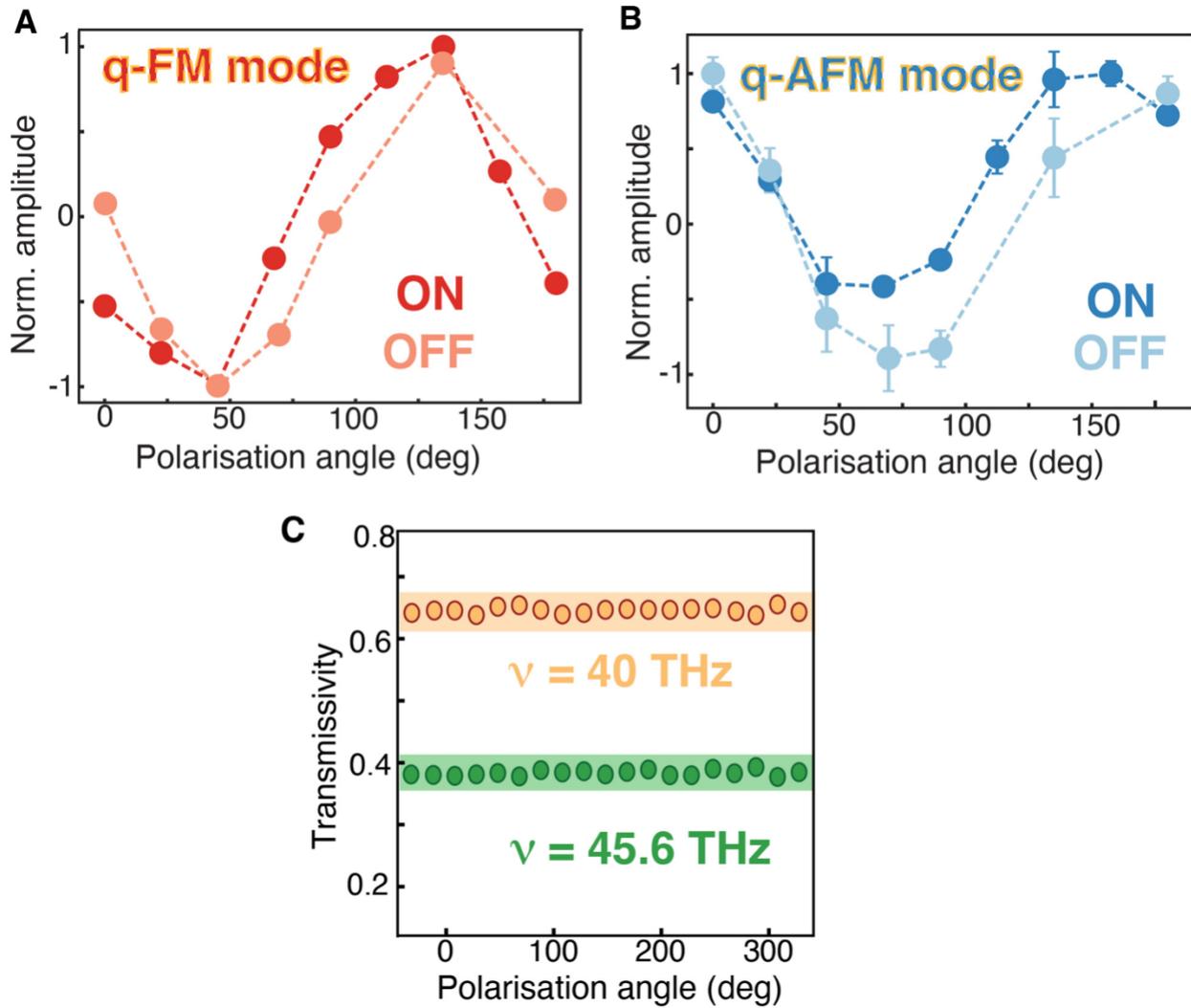

**Fig. S13. Pump polarization dependence of the q-FM mode, q-AFM mode and absorption.** The normalized amplitude of the q-FM mode (A) and q-AFM (B) mode depends 180°-periodically on the angle of linear polarization of the pump beam. The probe beam was linearly polarized along the direction orthogonal to the magnetic field. The fluence was set to 6 mJ/cm$^2$. (C) Polarization dependence of the absorption of hematite in the ab-plane for mid-IR frequencies resonant and off-resonant with the 2M mode. The green and yellow-shaded area corresponds to the experimental uncertainty.

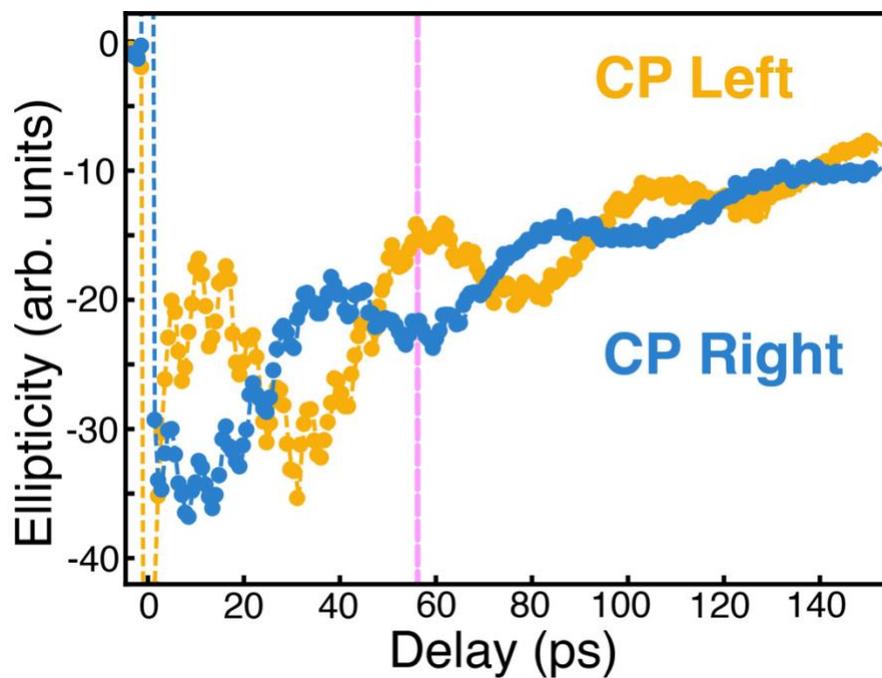

**Fig. S14. Spin dynamics induced by circularly polarized mid-infrared laser pulses.** The probe beam was linearly polarized along the direction orthogonal to the magnetic field. The fluence was set to 6 mJ/cm².